\documentstyle[epsfig,12pt]{article}
\begin{document}

\author{Vesselin G. Gueorguiev and Jerry P. Draayer \\
{\it {Louisiana State University, Baton Rouge, USA}}}
\title{{\bf {Basis Generator for M-Scheme SU(3) Shell Model Calculations}}}
\date{\today \thanks{
Paper published in: ``Revista Mexicana de Fisica'' {\bf 44 }
Suplemento {\bf 2, 43-47 (}October 1998{\bf )}}}

\maketitle

\begin{abstract}
A FORTRAN code for generating the leading $SU(3)$ irreducible representation
(irrep) of $N$ identical spin $^{1}/_{2}$ fermions in a harmonic oscillator
mean field is introduced. The basis states are labeled by $N$ -- the total
number of particles, the $SU(3)$ irrep labels ($\lambda ,\mu $), and $S$ --
the total spin of the system. The orthonormalized basis states have two
additional good quantum numbers: $\epsilon $ -- the eigenvalue of the
quadruple operator, $Q_{0}$; and $M_{J}$ -- the eigenvalues of the
projection of the total angular momentum operator, $J_{0}=L_{0}+S_{0}$. The
approach that is developed can be used for a description of nuclei in a
proton-neutron representation and is part of a larger program aimed at
integrating $SU(3)$ symmetry into the best of the currently available exact
shell-model technologies.
\end{abstract}

\section{Introduction}

Successful models for describing energy spectra, transition strengths and
other nuclear phenomena usually work well in one region of a shell but not
in others. For example, the random phase approximation (RPA) \cite{RPA} is a
reasonable theory for describing properties of nuclei near closed shells but
fails in mid-shell regions where deformation is the most characteristic
feature due to the importance of the quadruple-quadrupole ($Q \cdot Q$)
interaction \cite{Quadruple Interaction} in this domain. For near mid-shell
nuclei, algebraic models based on $SU(3)$ \cite{SU(3) Models} work best
since the basis states are then eigenstates of $Q \cdot Q$.

Applications of a nuclear shell-model theory always involve three
considerations \cite{Shell Model Procedures}:

\begin{itemize}
\item  {\em Step 1:} Selection of a model space based on a simplified
Hamiltonian. Frequently used approximations include the Simple Harmonic
Oscillator Hamiltonian for near closed-shell nuclei and a Nilsson
Hamiltonian for deformed systems.

\item  {\em Step 2:} Diagonalization of a realistic Hamiltonian in the model
space to obtain the energy spectrum and eigenstates of the system under
consideration. Important components of a realistic Hamiltonian include the 
$Q\cdot Q$ and pairing ($P$) interactions as well as single-particle 
$l\cdot l$ and $l\cdot s$ terms.

\item  {\em Step 3:} Evaluation of electromagnetic transition strengths (E2,
M1, etc.) between calculated eigenstates and a comparison with the available
experimental data.
\end{itemize}

\noindent If the model Hamiltonian includes free parameters, for example,
single-particle energies and/or the strength of the $Q \cdot Q$ and $P$
interactions, the procedure is repeated to obtain a best overall fit to the
experiment data.

The easiest to use of modern shell-model codes are based on the so-called M-scheme
logic, namely, model spaces spanned by many-particle configurations (Slater
determinants) with good projection of the total angular momentum \cite
{ANTOINE,OXBASH}. Good total angular momentum, which is a conserved symmetry
due to the isotropy of space, is obtained by angular momentum projection.
Codes of this type normally yield a good description of the single-particle
nuclear phenomena; unfortunately, an equally good description of collective
phenomena within the framework of such a theory is very difficult to
achieve. On other hand, an $SU(3)$ based shell-model scheme is designed to
give a simple interpretation of collective phenomena. An ideal scenario
would incorporate both, allowing the Hamiltonian of the system to ``choose''
which one of the two (or an admixture) is most appropriate.

A project, code named Red Stick for Baton Rouge, is now under development at
Louisiana State University. Principal authors are Jerry Draayer, Erich
Ormand, Calvin Johnson, and Vesselin Gueorguiev. The goal of the project is
to develop a new M-scheme shell-model code that allows the user to include 
$SU(3)$ symmetry-adapted basis states in addition to (the usual)
single-particle based configurations.

In this paper we discuss the first stage of the Red Stick project -- a basis
generator for $M$-scheme $SU(3)$ shell-model configurations. We begin with a
review of $SU(3)$ shell-model basics \cite{Elliott I, Elliott III}. Then we
introduce an appropriate single-particle level scheme and give matrix
elements of the generators of $SU(3)$ in that representation. Next we
consider the structure of the Highest Weight State (HWS) of an $SU(3)$
irrep, and especially the HWS of so-called leading $SU(3)$ irreps. Once a
HWS is known, we then go on to show how all states of that irrep can be
constructed using $SU(3)$ step operators \cite{Hecht}. States with good
projection of the total angular momentum $M_{J}$ are obtained by considering
the direct product $SU(3) \otimes SU_S(2)$. Some information about the
FORTRAN code that uses these techniques will also be presented.

\subsection{$SU(3)$ Basics}

In this section we review group theoretical concepts that are important to
the development of the theory and introduce $SU(3)$ conventions adopted in
our discussion. We also consider the physical reduction $SU(3) \supset SO(3)$
and the canonical reduction $SU(3)\supset U(1)\otimes SU(2)$ with their
respective labels.

First consider the so-called physical reduction $SU(3)\supset SO(3)$, which
yields a convenient labeling scheme for the generators of $SU(3)$ in terms
of $SO(3)$ tensor operators. The commutation relations for these $SU(3)
\supset SO(3)$ tensor operators are given in terms of ordinary $SO(3)$
Clebsch-Gordan coefficients (CGC) \cite{Elliott I}: 
\begin{eqnarray}
\lbrack L_{m},L_{m^{\prime }}] &=&-\sqrt{2}(1m,1m^{\prime }|1m+m^{\prime
})L_{m+m^{\prime }}  \nonumber \\
\lbrack Q_{m},L_{m^{\prime }}] &=&-\sqrt{6}(2m,1m^{\prime }|2m+m^{\prime
})Q_{m+m^{\prime }}  \label{LQ - Elliott I} \\
\lbrack Q_{m},Q_{m^{\prime }}] &=&3\sqrt{10}(2m,2m^{\prime }|1m+m^{\prime
})L_{m+m^{\prime }}  \nonumber
\end{eqnarray}
Within this reduction scheme, states of an $SU(3)$ irrep $(\lambda,\mu )$
have the following labels:

\begin{itemize}
\item  $(\lambda ,\mu )$ -- $SU(3)$ irrep labels;

\item  $l$ -- total orbital angular momentum, which corresponds to the second
order Casimir operator of $SO(3)$;

\item  $m_{l}$ -- projection of the angular momentum along the laboratory 
$z$-axis;

\item  $k$ -- projection of the angular momentum in a body-fixed frame, which
is related to multiple occurrences of $SO(3)$ irreps with angular momentum 
$l$ in the $(\lambda ,\mu )$ irrep.
\end{itemize}

\noindent Unfortunately, this scheme has only one additive label, namely 
$m_{l}$, and in addition, there are technical difficulties associated with
handling the $k$ label.

The labeling scheme for this project is the canonical reduction 
$SU(3) \supset U(1) \otimes SU(2)$ where $Q_{0}$ is the $U(1)$ generator and
the $SU(2)$ generators are proportional to $L_{0}$, $Q_{+2}$, and $Q_{-2}$ 
\cite{Elliott I}. Under the action of the generators of these $U(1)$ and 
$SU(2)$ groups, the remaining four generators of $SU(3)$ transform as two
conjugate spin $[\frac{1}{2}]$ $SU(2)$ tensors with $\varepsilon = \pm 3$
values for $Q_{0}$. For this scheme, states of a given $SU(3)$ irrep
$(\lambda,\mu )$ have the following labels:

\begin{itemize}
\item  $(\lambda ,\mu )$ -- $SU(3)$ irrep labels,

\item  $\varepsilon $ -- quadruple moment ($Q_{0}$),

\item  $m_{l}$ -- projection of the orbital angular momentum ($L_{0}$),

\item  $n_{\rho }$ -- related to the second order Casimir operator of $SU(2)
$, which for symmetric $(\lambda ,0)$ irreps is simply the number of
oscillator quanta in the $(x,y)$ plane.
\end{itemize}

This canonical reduction, $SU(3)\supset U(1)\otimes SU(2)$, has two additive
labels, $\varepsilon $ ($Q_{0}$) and $m_{l}$ ($L_{0}$) and the allowed
values of these labels for fixed $SU(3)$ irrep $(\lambda ,\mu )$ are given
by \cite{Hecht}:

\begin{eqnarray}
\varepsilon  &=&2\lambda +\mu -3(p+q)  \label{pqm-parametriztion} \\
n_{\rho } &=&\mu +(p-q)  \nonumber \\
m_{l} &=&n_{\rho }-2m  \nonumber
\end{eqnarray}

where $0\leq p\leq \lambda $ , $0\leq q\leq \mu $ , and $0\leq m\leq n_{\rho
}.$

\begin{figure}[h]
\centerline{\hbox{\epsfig{figure=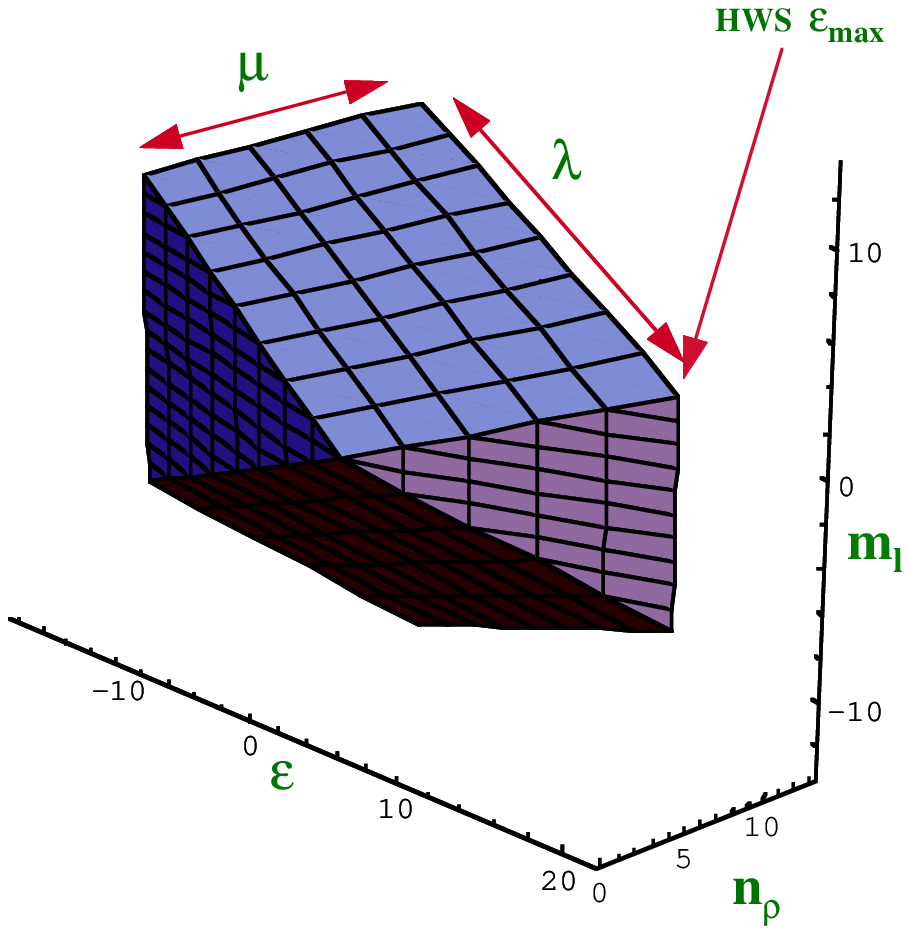,width=8cm,height=8cm}}}
\caption{Three-dimensional view of the $(\lambda ,\mu )$ $SU(3)$ irrep.}
\label{3D-view of SU(3) irrep}
\end{figure}

\section{Basis State Generation}

The algorithm we use to generate $SU(3)$ symmetry adapted states with good
projection of the total angular momentum is described in this section. It
consists of four basic components: 1) definition of the single-particle
levels and matrix elements of the $SU(3)$ generators for a given shell; 2)
generation of the HWS of $SU(3) \otimes SU_{S}(2)$
for given number of fermions $N$ and spin $S$; 3) generation of the other 
$SU(3)$ states by applying step operators to the HWS; and 4) generation of
states of good total angular momentum projection, $M_{J}$. This will be
followed by a short discussion of a FORTRAN code that uses the algorithm.

\subsection{Single-particle Levels -- Ordering Scheme}

\begin{figure}[h]
\centerline{\hbox{\epsfig{figure=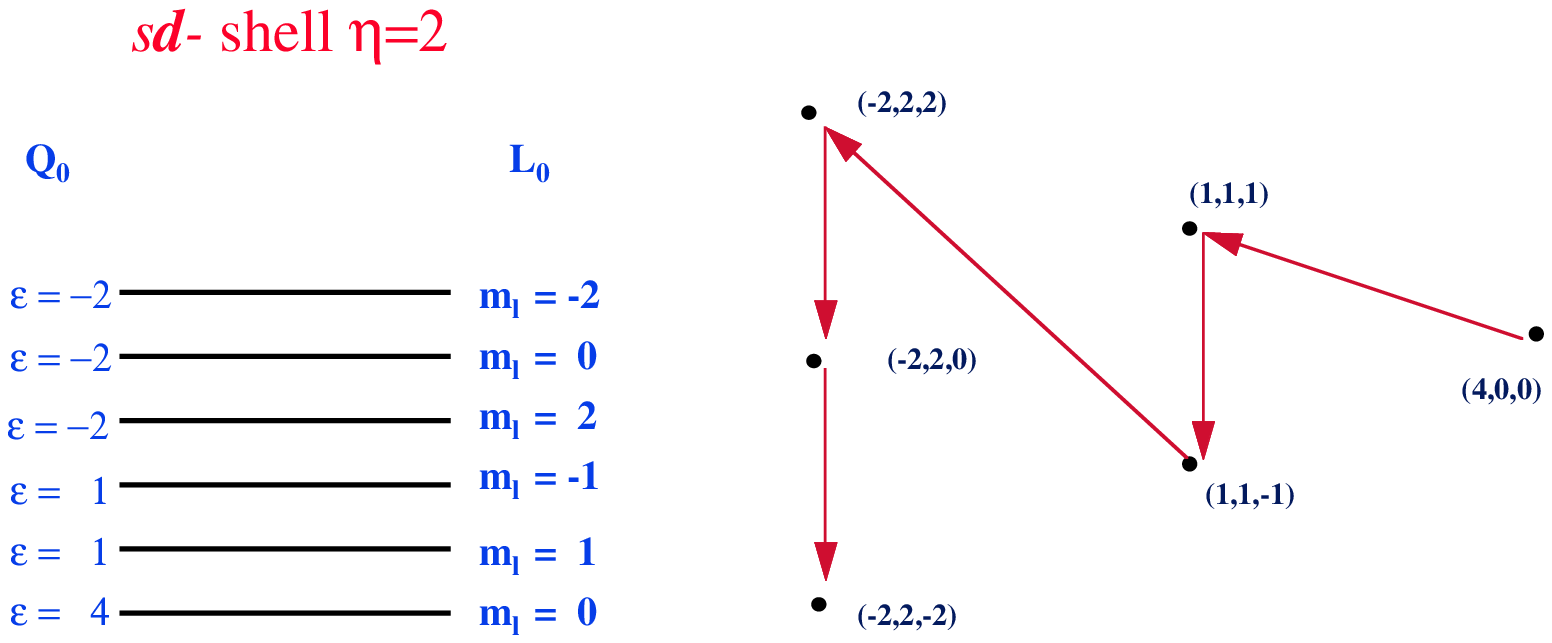,width=11cm,height=6cm}}}
\caption{Ordering scheme of the single-particle levels.}
\label{Single Particle Levels}
\end{figure}

Single-particle levels of the $\eta = 0,1,2,...$ $(s,p,sd,...)$ harmonic
oscillator shell belong to the symmetric $(\eta,0)$ irrep of $SU(3)$.
Because $\mu = 0$, a typical three-dimensional representation of $SU(3)$ basis
states, Fig.\ref{3D-view of SU(3) irrep}, reduces to a special
two-dimensional triangular shape ($\varepsilon$ and $n_{\rho}$ become
linearly dependent), Fig.\ref{Single Particle Levels}. Also, because $SU(3)$
is a compact group its irreps are finite dimensional and many-particle
(fermion) configurations can be conveniently represented as binary strings
with a $1$ or $0$ representing the presence or absence of a particle in the
corresponding single-particle level. (The latter, together with a ``sign
rule'' to accommodate fermion statistics, is a convenient computer native
implementation of a Slater determinant representation of the basis states.)

A convenient ordering scheme (which tracks the arrows in Fig.\ref{Single
Particle Levels}) is set by requiring a simple representation of
many-particle configurations with maximum quadruple deformation. This
objective can be achieved if the states are ordered by $\varepsilon$
(quadruple moment) first and then by $m_{l}$ (projection of the total
angular momentum).

\subsection{Action of $SU(3)$ Generators on Single-particle States}

\begin{figure}[h]
\centerline{\hbox{\epsfig{figure=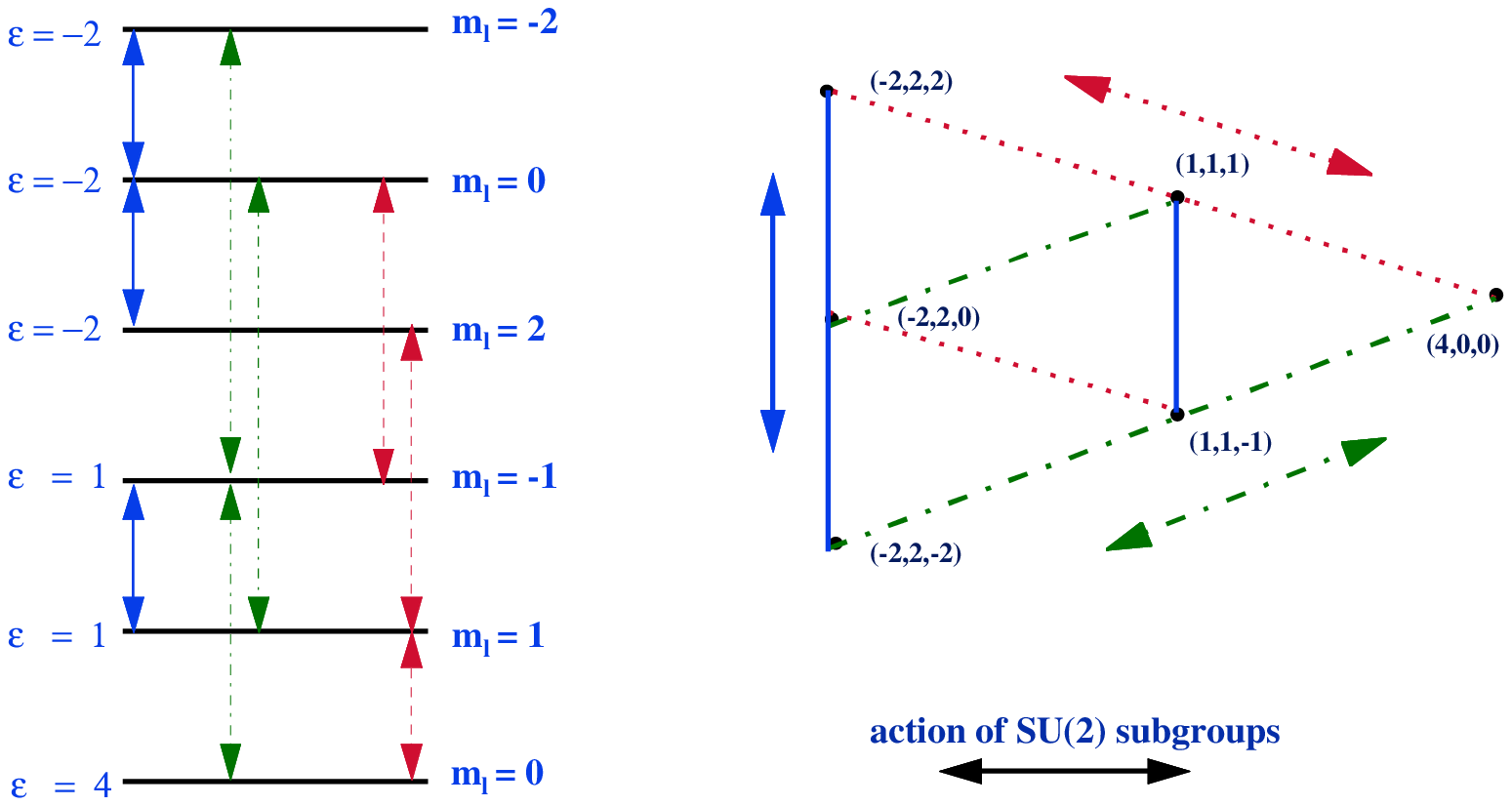,width=11cm,height=6cm}}}
\caption{The diagram on the right shows the action of the six non-diagonal
generators of $SU(3)$. The vertical solid lines represent the action of the
$SU(2)$ subgroup that enters the $SU(3) \supset SU(2)\otimes U(1)$ chain. The
diagram on the left shows that applying an $SU(3)$ generator to  a
single-particle state results in another single-particle state.}
\label{SU(3) Matrix Elements}
\end{figure}

To be able to apply the $SU(3)$ generators on many-particle configurations
it suffices to know the action of these generators on the single-particle
states. The eight generators of $SU(3)$ belong to the self-adjoint $(1,1)$
irrep of $SU(3)$. The operator structure should be chosen in the most
convenient form for the application under consideration. For the present
application, this choice is the same as used for the basis states, namely,
the  $SU(3)\supset SU(2)\otimes U(1)$ reduction. The matrix elements of the
$SU(3)$ generators can be obtained either by using an application of the
appropriate Wigner-Eckert theorem or by using explicit expressions
\cite{Hecht} for determining the action of the operators on the basis states.
For computational purposes it is better to adopt  a direct solution, one that
exploits the fact that the action is on a product of single-particle levels
each of which belongs to a symmetric $(\eta ,0)$ irrep of $SU(3)$. This
allows the matrix elements of the $SU(3)$ generators to be calculated using
properties of the $SU(2)$ only (Fig.\ref{SU(3) Matrix Elements}).

A key feature is the fact that the six non-diagonal generators of $SU(3)$
(recall that $L_{0}$ and $Q_{0}$ are diagonal) are rising or lowering
generators of $SU(2)$ subgroups of $SU(3)$. The three $SU(2)$ subgroups and
their respective actions are shown in Fig. \ref{SU(3) Matrix Elements}. States
that are collinear with one of the sides of the triangular shape shown on the
right in the figure form an irrep of the corresponding $SU(2)$ subgroup.

\subsection{Highest Weight States of $SU(3) \otimes SU_{S}(2)$ for Leading
Irreps}

So far we have constructed single-particle states and evaluated matrix
elements of the generators of $SU(3)$ when they act on these states. The
next step is to construct many-particle HWS of $SU(3) \otimes
SU_{S}(2)$. In the chosen scheme there are seven extreme states which correspond
to the vertexes of the three-dimensional diagram (Fig.\ref{3D-view of SU(3) irrep}) of a
general $(\lambda,\mu)$ irrep. We are particularly interested in the vertex that has
the maximum value for the quadruple moment of the system (Fig.\ref{3D-view of SU(3)
irrep}). Our HWS is the state with $\varepsilon = 2\lambda +\mu$, $n_{\rho }=\mu$, and
$m_{l}=\mu$. This HWS (maximum value of $m_{l}$ for maximum $\varepsilon$) can be easily
constructed by ensuring that the action of the $SU(3)$ rising generators  annihilate it.
Indeed, for such  a HWS the values of $\lambda$ and $\mu $ can be determined from its
$\varepsilon$ and $m_{l}$ labels.

Selecting the leading $(\lambda,\mu)$ irrep (HWS with maximum overall value
of $\varepsilon $) out of all possible irreps of an $N$ fermion system with
total system spin $S$ is very simple within the chosen scheme. This is
because the number of particles with spin up $n_{\uparrow }$ and spin down 
$n_{\downarrow }$ is uniquely determined by the solution of two linear
equations: 
\begin{eqnarray*}
N &=&n_{\uparrow }+n_{\downarrow } \\
2S &=&n_{\uparrow }-n_{\downarrow }.
\end{eqnarray*}
Where the second equation expresses the fact that we also require the state
to be highest weight with respect to $SU_{S}(2)$. Further, maximizing the
value of $Q_{0}$ is achieved by filling the single-particle states of the 
$(\eta ,0)$ irrep (Fig.\ref{Single Particle Levels}) from bottom to top. The
chosen scheme ensures that this simple procedure gives maximum values for 
$\varepsilon$ and $m_{l}$. The $SU(3)$ irrep labels $(\lambda,\mu)$ are
obtained by evaluating the quadruple moment $Q_{0}$ and projection of the
angular momentum $L_{0}$, as these are additive quantum numbers.

\begin{figure}[h]
\centerline{\hbox{\epsfig{figure=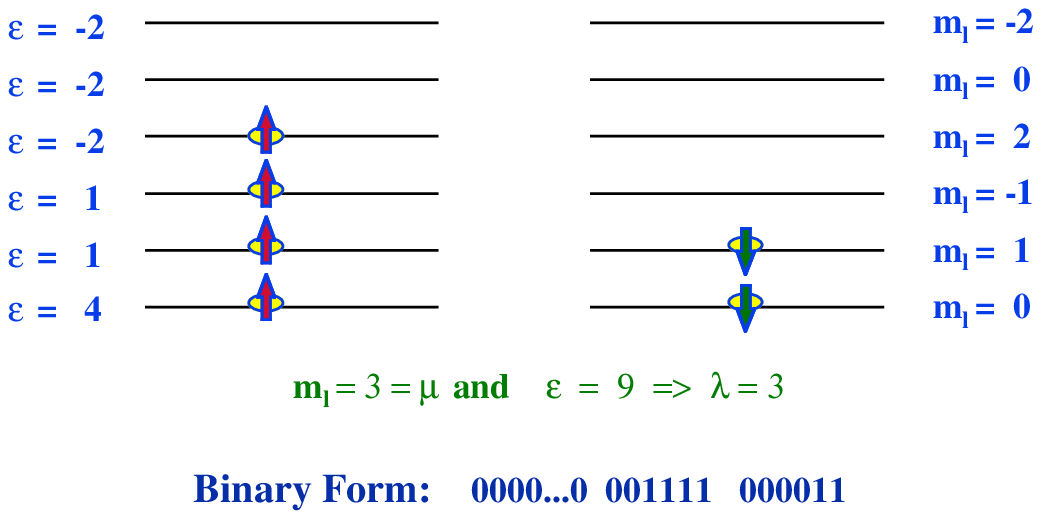,width=10cm,height=5.5cm}}}
\caption{Highest weight state of the leading $(3,3)$ irrep for $N=6$ and
$S=1$ in the $sd$--shell.}
\label{N=6&S=1}
\end{figure}

\begin{figure}[h]
\centerline{\hbox{\epsfig{figure=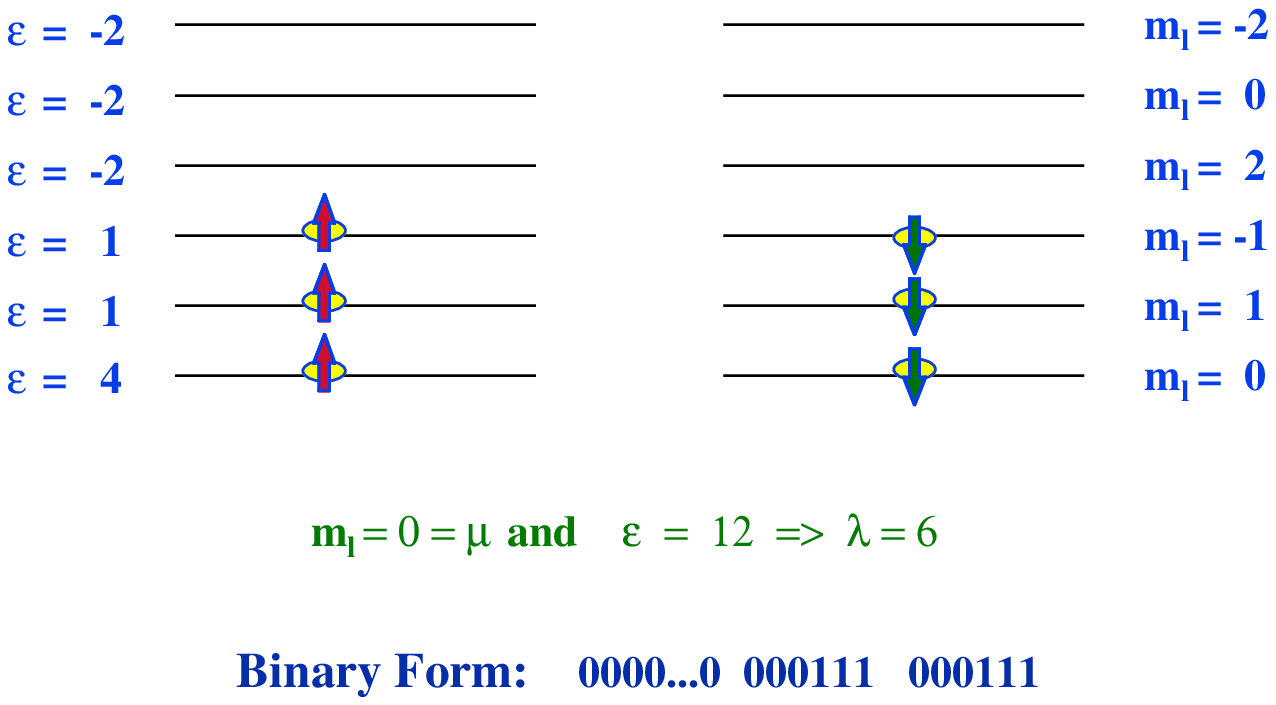,width=12cm,height=5.5cm}}}
\caption{Highest weight state of the leading $(6,0)$ irrep for $N=6$ and $S=0$
in the $sd$--shell. }
\label{N=6&S=0}
\end{figure}

For example, in the $sd$--shell, there are six single-particle levels
corresponding to the $(2,0)$ irrep of $SU(3)$. The HWS of the leading irrep for $N=6$
particles and total system spin $S=1$ is $(3,3)$ (Fig.\ref{N=6&S=1}), whereas for $S=0$
the leading $SU(3)$ irrep is $(6,0)$ (Fig.\ref{N=6&S=0}). These many-particle
configurations are HWS with respect to $SU(3)$ as well as $SU_{S}(2)$.

\subsection{Generating $SU(3)$ States with Step Operators}

\begin{figure}[h]
\centerline{\hbox{\epsfig{figure=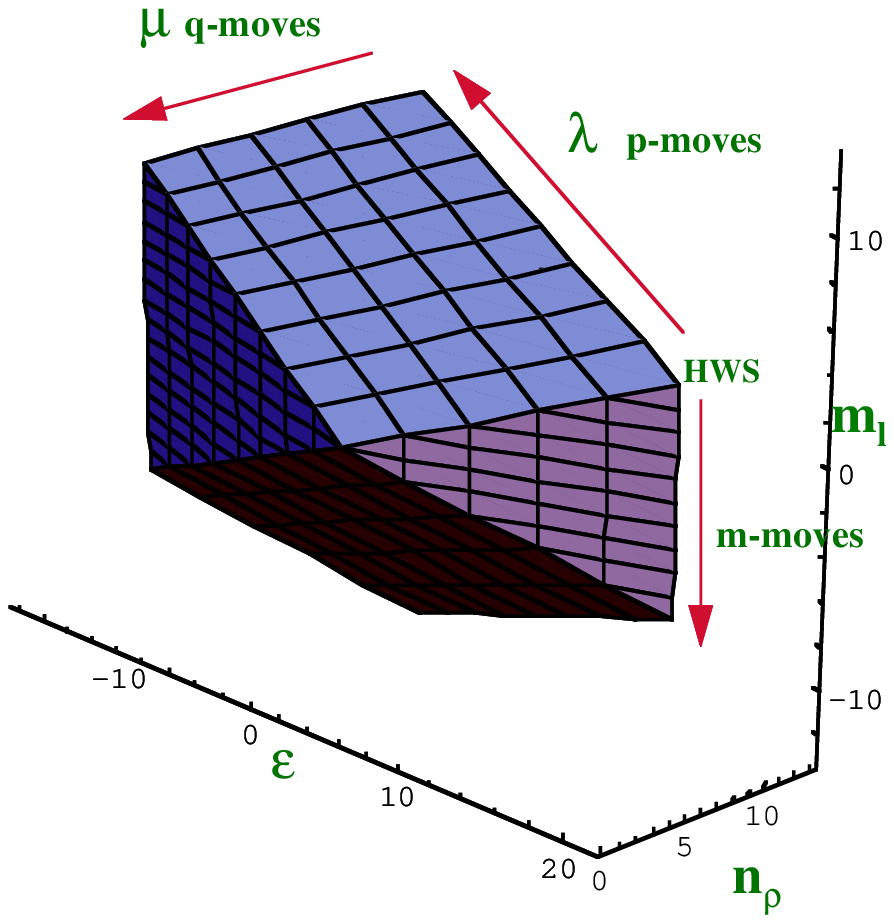,width=9cm,height=9cm}}}
\caption{Action of the $SU(3)$ step operators.}
\label{SU(3) Step Operators}
\end{figure}

Once we have the HWS of $SU(3)$ $\otimes $ $SU_{S}(2)$, we can generate any
other state of the $SU(3)$ irrep $(\lambda ,\mu )$ by applying step
operators similar to those given by Hecht. By using the parameterization (\ref
{pqm-parametriztion}) \cite{Hecht}, we can identify the corresponding step
operators, Fig. \ref{SU(3) Step Operators}. 

It is important to note that applying $p$--move or $q$--move step operators to
states on the top surface yields other states (or zero) on that same
surface. Since the states on the top surface are HWS with respect to $SU(2)$
in the $SU(3)$ $\supset $ $SU(2)\otimes U(1)$ reduction, the $m$--move step
operator is an $SU(2)$ lowering operator which changes the projection of the
angular momentum, $m_{l}$. The $p$--move and $q$--move step operators can be
obtained by imposing the restriction that they generate only transformations
within the $SU(2)$ HWS space. From an algebraic perspective, the $p$--move
and the $m$--move operators are linear in $SU(3)$ generators while the $q$--move
operator is quadratic. Nevertheless, the state generation process can be written in such
a way that the $q$--move operator effectively reduces to a linear action. These step
operators can also be obtained by a projection operator technique \cite{Tolstoy}.

\subsection{Generating States of Good $M_{J}$ by Lowering the Spin}

Since the action of $SU(3)$ commutes with that of the spin, $SU_{S}(2)$, it
is not difficult to achieve the final goal of states with good total angular
momentum projection, $M_{J}$. Recall that we started with the HWS of $SU(3)$ 
$\otimes $ $SU_{S}(2)$. We then introduced a procedure to generate other
states of the $SU(3)$ irrep by applying step operators. Each of these states
remains a HWS of $SU_{S}(2)$. By applying spin lowering operators to $SU(3)$
states with $m_{l}=M_{J}-m_{s}$, we can generate states with labels: $N$, $S$
, $(\lambda,\mu)$, $\varepsilon$, $n_{\rho}$, $m_{l}$, and $M_{J}$.

\subsection{About the FORTRAN code}

A FORTRAN code that uses the above algorithm was run on three platforms:
Macintosh IIci, Power Macintosh 8500, and SUN workstation. Here we give some
of the run-time characteristics of the code for shells below the $sdg$--shell:

\begin{itemize}
\item  On a $50$ MHz Macintosh IIci with math co-processor, $3$ Mb of memory
suffices and the longest run time for generating all states of any leading 
$SU(3)$ irrep was less than $5$ min.

\item  On a $200$ MHz Power Macintosh $8500$, $4$ Mb of memory was
sufficient and the longest run time for generating all states of any leading 
$SU(3)$ irrep was less than $30$ sec.

\item  On a $200$ MHz SUN workstation the longest run time for generating
all states of any leading $SU(3)$ irrep was less than $15$ sec. All states
with good $M_{J}$ for any leading irrep of $SU(3)$ $\otimes $ $SU_{S}(2)$
for the $p,sd,$ and $pf$ shells can be generated in less than $20$ min.
\end{itemize}

Beyond the $pf$--shell ($\eta \geq 4$), the number of configurations in a
single $SU(3)$ state may exceed $100,000$. The storage requirement must be
pushed up accordingly. More storage is also needed because beyond the $sdg$--shell the
number of single-particle states exceed $16$ ($32$ with the spin degree of freedom) which
means more than a single word is required to represent many-particle configurations.

\section{Summary and Discussion}

An $SU(3)$ basis state generator for producing states with good projection
of the total angular momentum, $M_J$, and therefore one that is suitable for
integration into modern $M$-scheme shell-model codes, is suggested as a
possible way to obtain simultaneously a description of collective and
single-particle nuclear phenomena. Indeed, a basis state generator that uses
the algorithm described above and generates $SU(3)$ states as a sum of
Slater determinants (many-particle configurations) with good $M_{J}$ is
available. The basis state generation procedure consists of four main
components as follows:

\begin{itemize}
\item  Symmetric irreps of $SU(3)$ are chosen as single-particle shell-model
states,

\item  many-particle state is constructed as HWS of $SU(3)$ 
$\otimes $ $SU_{S}(2)$,

\item  $SU(3)$ step operators are applied to obtain other states within an 
$SU(3)$ irrep,

\item  and finally, states with good $M_{J}$ are then obtained by $SU_{S}(2)$
lowering operators.
\end{itemize}

There are additional things that need to be done, especially the generation
of other then leading irreps of $SU(3)$. This can be achieved by employing
an algorithm for generating non-leading HWS that is part of an $SU(3)$
reduced matrix element package \cite{Bahri} or perhaps by using another more
direct approach tailored to the present application. The generation of other
HWS is important if one wishes to include states that are not maximally
deformed in their intrinsic configuration in a calculation. For example,
these will be important if the non-$Q \cdot Q$ parts of the interaction play
a significant role.

The generation of good proton-neutron $SU(3)$ and $M_{J}$ states can be
achieved in a variety of ways. One scheme, the so-called strong coupling
limit uses standard $SU(3)$ coupling procedures to couple the proton and
neutron irreps to final irreps: proton states, $|N^{\pi}$ $S^{\pi}$ 
$(\lambda^{\pi},\mu^{\pi})$ 
$\varepsilon^{\pi},n_{\rho}^{\pi},m_{l}^{\pi},M_{J}^{\pi}>$,
are coupled with neutron states, $|N^{\nu}$ $S^{\nu}$ $(\lambda^{\nu},\mu^{\nu})$
$\varepsilon^{\nu},n_{\rho}^{\nu},m_{l}^{\nu},M_{J}^{\nu}>$, 
to final states $|N$ $S$ $(\lambda ,\mu )$ $\varepsilon
,n_{\rho },m_{l},M_{J}>$. Another procedure would be to extend what was done
here for $SU_S(2)$ to $SU(4)$ reduced with respect to $SU_S(2)$ $\otimes$ 
$SU_T(2)$, the supermultiplet scheme. The latter is not simple because the 
$SU(4)$ HWS is not the one of primary interest in nuclear physics and
therefore it must be reduced, which requires procedures like those developed
here for a non-canonical reduction of $SU(4)$.

The underlying goal is to be able to carry out shell-model calculations
using $M$-scheme techniques with $SU(3)$ symmetry adapted basis states
included in the basis. The $SU(3)$ basis state generator described here
provides good start in that direction, and therefore the beginning of an
symmetry adapted $M$-scheme code that is an integral part of our Red Stick
shell-model project.

\end{document}